\documentclass[aps,prb,twocolumn,floatfix,showpacs,superscriptaddress,nobibnotes]{revtex4-1}

\usepackage{amsmath,amsfonts}
\usepackage{graphicx}

\usepackage{dcolumn}
\usepackage{longtable}
\usepackage{color}
\usepackage{setspace}

\def\Journal #1,#2,#3,#4#5#6#7{#1 {\bf #2}, #3 (#4#5#6#7)}

\def\gsim{\lower -0.3ex \hbox{$>$} \kern -0.75em \lower 0.7ex
	\hbox{$\sim$}}
\def\lsim{\lower -0.3ex \hbox{$<$} \kern -0.75em \lower 0.7ex
	\hbox{$\sim$}}

\def\Vec#1{{\bf #1}}
\def\GVec#1{\mbox{\boldmath $#1$}}

\begin{document}

\title{Incommensurate double-walled carbon nanotubes as one-dimensional moir\'e crystals}

\author{Mikito Koshino}
\thanks{All authors contributed to the manuscript extensively.}
\affiliation{Department of Physics, Tohoku University, Sendai, 980--8578, Japan}

\author{Pilkyung Moon}
\thanks{All authors contributed to the manuscript extensively.}
\affiliation{New York University Shanghai, Pudong, Shanghai 200120, China}

\author{Young-Woo Son}
\thanks{All authors contributed to the manuscript extensively.}
\affiliation{Korea Institute for Advanced Study, Seoul 130-722, Korea}

\begin{abstract}
Cylindrical multishell structure is one of the prevalent atomic arrangements in nanowires. 
Being multishell, the well-defined atomic periodicity is hardly realized in it
because the periodic units of individual shells therein 
generally do not match except for very few cases, 
posing a challenge to understand its physical properties. 
Here we show that moir\'e patterns generated by superimposing atomic lattices 
of individual shells are decisive in determining its electronic structures. 
Double-walled carbon nanotubes, as an example, are shown to have spectacular 
variations in their electronic properties from metallic to semiconducting and 
further to insulating states depending on their moir\'e patterns, 
even when they are composed of only semiconducting nanotubes 
with almost similar energy gaps and diameters. 
Thus, aperiodic multishell nanowires can be classified 
into new one-dimensional moir\'e crystals with distinct electronic structures.
\end{abstract}

\maketitle

\section{Introduction}
When repetitive structures are overlaid against each other, 
a new superimposed moir\'e pattern emerges and is observed in various macroscopic phenomena~\cite{oster}.
Recent progress in stacking two-dimensional crystals~\cite{geim} enables 
the patterns to occur at the atomic scale, 
showing their distinct quantum effects~\cite{yankowitz,ponomarenko,dean,hunt}.
Even in one-dimension, this atomic pattern is realized naturally
in the multishell organic and inorganic tubular shaped nanowires~\cite{iijima,ge,tenne}. 
Among them, the double-walled carbon nanotubes (DWNTs) formed by
two concentric single-walled carbon nanotubes (SWNTs)
are the simplest multi-shell nanotube structures~\cite{saito}. 
%
The electronic structure of SWNT, a basic building block of DWNTs,
depends on its way of rolling a single layer graphene
along a specific chiral vector into a seamless cylindrical shape. 
The chiral vector, ${\bf C}=n{\bf a}_1 +m{\bf a}_2$, or a set 
of integers $(n,m)$ uniquely determines electronic structures of SWNTs where
${\bf a}_1$ and ${\bf a}_2$ are the primitive vectors of hexagonal lattice of graphene
[Fig.\ 1(d)].
They are metallic if $|n-m|$ is a multiple of three 
and otherwise semiconducting~\cite{saito,ando,charlier_review}.
This simple rule can be obtained by reducing or quantizing one dimension
of the two dimensional massless Dirac energy bands of graphene.

In spite of such a clear rule, 
its extension to double-walled 
structures is far from trivial~\cite{shen,ando,charlier_review}. 
Ever since its discovery~\cite{iijima}, 
direct {\it ab initio} or empirical calculations have been performed to obtain the 
energy bands of DWNTs only if two single-walled nanotubes have a common periodicity
along its axis~\cite{saito,ando,charlier_review,shen,saito2,charlier,okada}.  
Very few of DWNTs, however, have commensurate atomic 
structures and the rest of them do not have the well-defined periodicity, posing
a significant challenge to understand their electronic properties. 
This situation also holds for other inorganic one-dimensional multishell tubular structures with 
several different atomic elements~\cite{tenne}.
In the literature, incommensurate DWNTs were studied numerically
in terms of the electronic structure \cite{lambin2000electronic,ahn}
and the transport properties,
\cite{roche2001conduction,yoon2002quantum,uryu2004electronic,uryu2005electronic}
and it is generally believed that the interlayer coupling does not strongly modify
the energy bands of the individual SWNTs.


On the other hand, there has been a rapid progress 
in stacking various two-dimensional crystals and 
in understanding their electronic properties~\cite{geim}. 
The most notable example among them is twisted bilayer graphene (TBLG)
where a single layer graphene is 
overlaid on top of the other with a rotational stacking fault~\cite{lopes,latil,hass,shallcross,kindermann}.  
These bilayer structures exhibit moir\'e 
patterns of which periodicity is quite larger than that of the unit cell of graphene.
When one layer rotates with respect to the other from zero to 60 degrees continuously, 
two hexagonal lattices can have a common exact supercell only for a few discrete rotation angles
while they cannot have the well-defined periodic 
unit for infinite possible other choices of angles \cite{shallcross,macdonald,falko,moon2}.

Formation of moir\'e pattern in TBLGs, however, do not require an exact matching of
atomic positions between the two layers for the common supercell, 
and its periodicity continuously changes as the angle 
varies~\cite{shallcross,macdonald,falko,moon2}.  
Recent theoretical~\cite{moon2} and 
experiment~\cite{havener} studies demonstrates 
that the electronic structure of TBLG is dictated 
not by the exactly matched atomic supercell but by the periodicity of moir\'e superlattice.
Therefore, successful descriptions of the electronic structures of TBLGs without 
commensurability validate the effective theory~\cite{macdonald,falko,moon2,moon3}
based on the Bloch wave expansion with 
respect to the moir\'e lattice in momentum space. 
This motivates us to explore a possible dimensional reduction from TBLGs with moir\'e patterns to one-dimensional structures 
which can be mapped onto DWNTs exactly. 
Using the effective theory and atomic structure mapping, we uncover 
that the moir\'e pattern plays a decisive role in determining electronic structures of DWNTs 
without any commensurability and that the resulting properties are
far beyond a simple sum of electronic bands
of two constituent nanotubes.

\section{Mapping from BLG to DWNT}

\begin{figure*}
	\begin{center}
			\leavevmode\includegraphics[width=0.8\hsize]{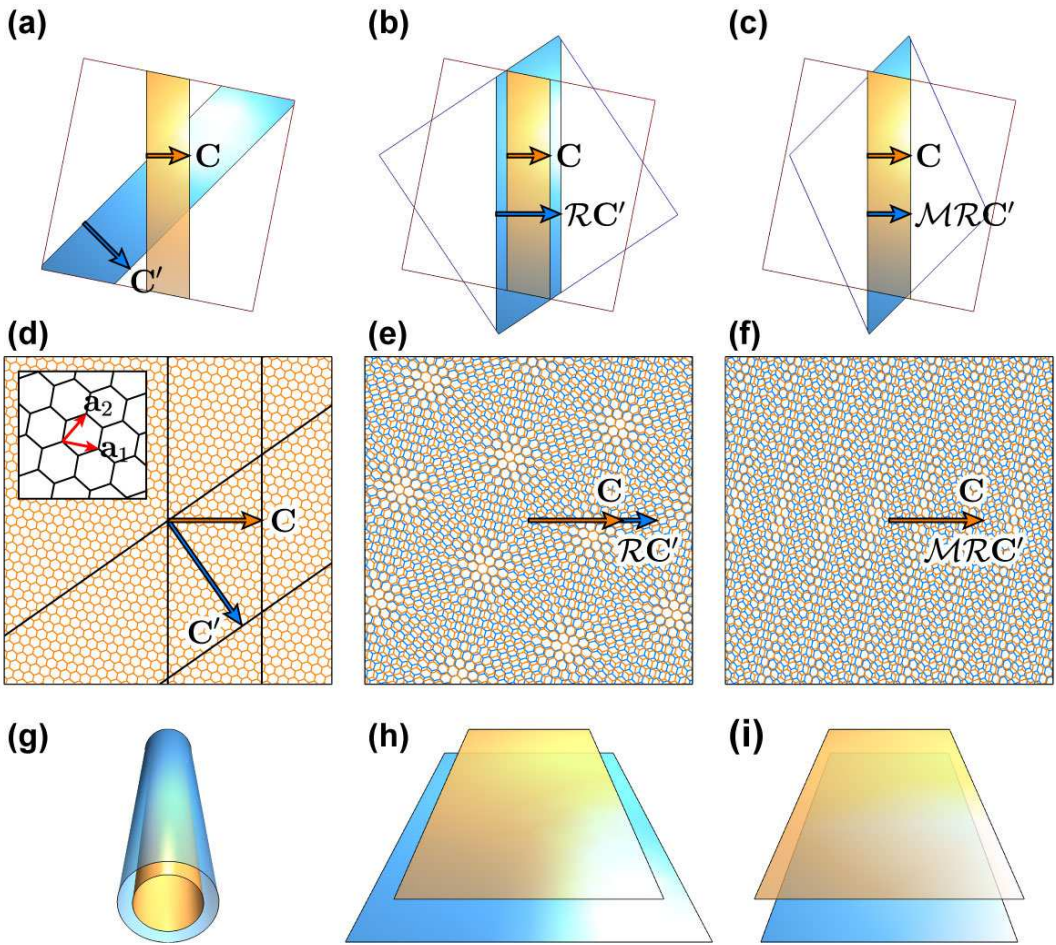}
	\end{center}
	\caption{
	Step by step operations for atomic structure mapping from graphene bilayer structure 
	to DWNT. 
	(a) Two parallelograms on upper and lower layers are drawn for unfolded 
	single-walled carbon nanotubes (SWNTs) 
	with different chiral vectors ${\bf C}$ and ${\bf C'}$ respectively. 
	(b) Rotation (${\mathcal R}$) and 
	(c) subsequent contraction (${\mathcal M}$) of lower layer
	to align the axial direction of two tubes and then to match their widths.
	A specific example for operations shown in (b) and (c)
	are displayed in (d), (e) and (f). 
	Here we use ${\bf C}=8{\bf a}_1+2{\bf a}_2$ and ${\bf C'}=14{\bf a}_1-10{\bf a}_2$
	for illustration
	where the common in-plane primitive vectors 
        ${\bf a}_1 = a(1,0)$ and ${\bf a}_2 = a(1/2,\sqrt{3}/2)$ 
        with the lattice constant $a\approx 0.246\, \mathrm{nm}$  
        are used to label the atomic positions of both layers. 
         In (d), thin solid lines perpendicular to chiral vectors 
	corresponds to parallelograms in (a).
	A usual two dimensional moir\'e pattern of TBLG is shown in (e) 
	and distorted TBLG with typical moir\'e lattice for DWNT is in (f). 
	Reverse mapping operations from a DWNT to double layer graphene nanoribbon
	are shown from (g) to (i). Unfolding (h)
	and then subsequent contraction of lower nanoribbon (i) map onto
	the modified TBLG structure [(c)and (f)] exactly.
	}
	\label{fig_schematics.ps}
\end{figure*}

We begin by describing atomic structure mapping procedures 
from bilayer graphene (BLG) to DWNT. 
This involves a rotation (its operator form is $\mathcal R$) 
and a subsequent uniaxial contraction ($\mathcal M$) of one layer with respect 
to the other in BLG.
The upper layer is designated for the inner tube 
with the chiral vector of ${\bf C}=n_1 {\bf a}_1 +n_2 {\bf a}_2$ 
and the lower for the outer with ${\bf C}'=n'_1{\bf a}_1 +n'_2{\bf a}_2$ [Figs.\ 1(a) and 1(d)].
First, the two different chiral vectors for the inner and outer SWNTs 
are aligned by rotating the lower layer, 
resulting in a usual TBLG with a moir\'e pattern [Figs.\ 1(b) and 1(e)]. 
After then, the lower one shrinks uniaxially along ${\bf C}$ 
to match the two chiral vectors exactly [Figs.\ 1(c) and 1(f)]. 
Resulting new primitive vectors for the lower layer become $\tilde{\bf a}_i={\mathcal {MR}}{\bf a}_i$
$(i=1,2)$. 
Corresponding reciprocal lattice vectors ${\bf b}_i$ and $\tilde{\bf b}_i$ for the upper and lower layers
can be defined to satisfy ${\bf a}_i \cdot {\bf b}_j =\tilde{\bf a}_i \cdot \tilde{\bf b}_j =2\pi\delta_{ij}$ $(i,j=1,2)$.
The exactly same atomic structure can be obtained by unfolding a DWNT 
into a bilayer graphene nanoribbon
and by subsequently shrinking the width of outer ribbon down to the inner one [Figs. 1(g) to 1(i)].
Therefore, the modified TBLG structure matches the atomic structure of DWNT with a periodic
boundary condition along $\bf C$ as shown in Fig.\ 1. 

\begin{figure*}
	\begin{center}
		\leavevmode\includegraphics[width=1.\hsize]{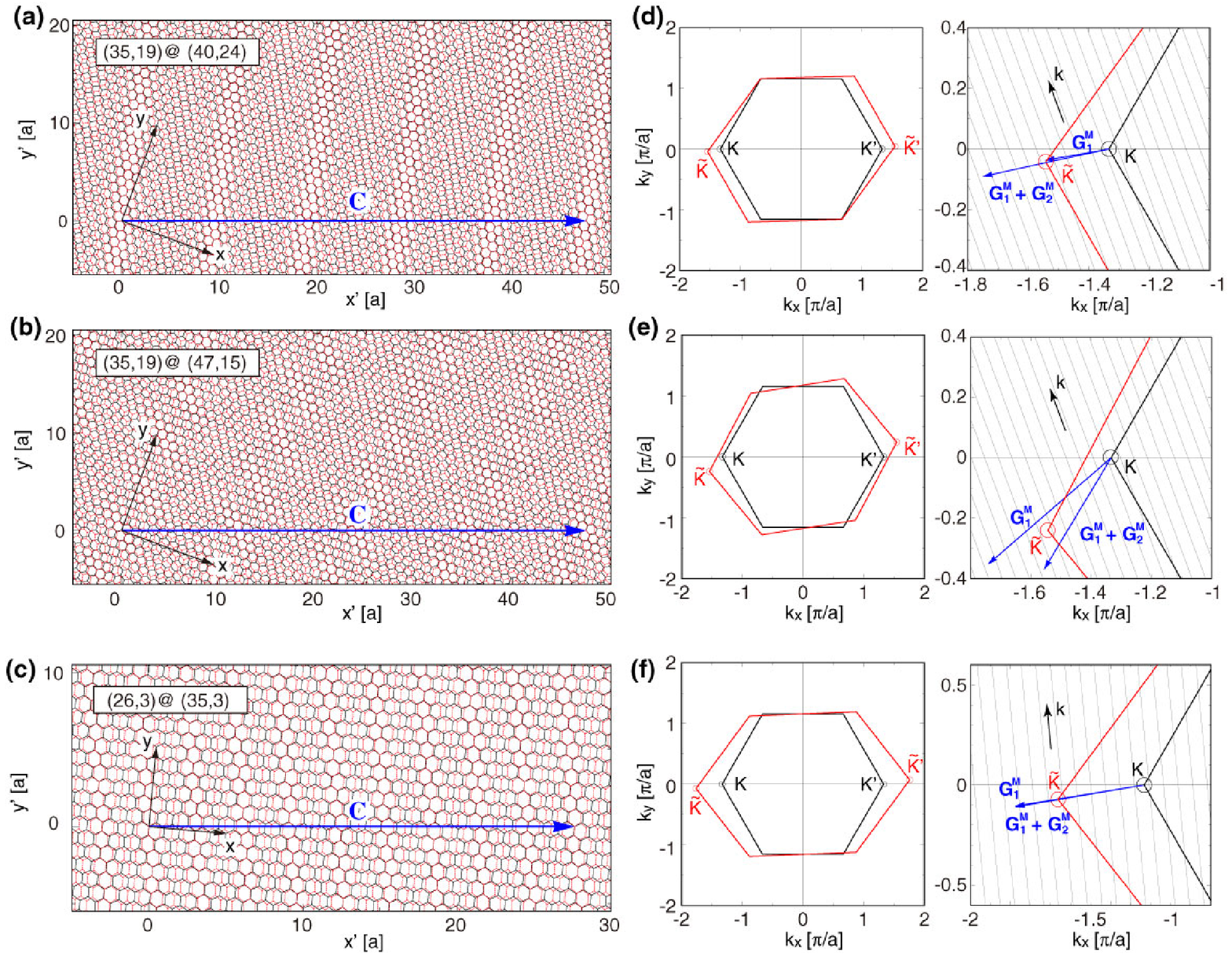}
	\end{center}
	\caption{
		After mapping operations in Fig. 1, two-dimensional atomic lattices corresponding 
		to (35,19) @(40,24), (35,19)@(47,15) and (26,3)@(35,3) DWNTs are drawn
		in (a), (b) and (c), respectively.
		Blue arrows are chiral vectors ({\bf C}) for each one.
		Normal (black) and distorted (red) hexagonal Brillouin zones (BZs) for three examples are 
		presented in (d), (e) and (f) respectively. 
		Upper (lower) Dirac points are indicated by ${\bf K}$ ($\tilde{\bf K}$). 
		Enlarged BZs near around its corner are shown in the right panels.
		In enlarged panels, thin slant lines are one-dimensional
		BZs separated by $2\pi/|{\bf C}|$ and ${\bf G}_\text{i}^\text{M}$ ($i=1,2$) is
		a reciprocal lattice vector corresponding to distorted moir\'e lattice.
	}
	\label{fig_lattice_and_bz.ps}
\end{figure*}

The mismatch between lattice periods of the upper and lower layers in the modified TBLG 
gives rise to the moir\'e superlattice pattern [Figs. 2(a)-2(c)]. 
In this structure, the arbitrary position $\bf r$ in the lower layer is displaced by 
$\boldsymbol{\delta} ({\bf r})=({\mathcal I}-{\mathcal R}^{-1}{\mathcal M}^{-1}){\bf r}$ by the mapping
where $\mathcal I$ is an identity operator.
The periodic vectors (${\bf L}^{\text M}_i$) of emerged moir\'e pattern can be obtained 
by using a condition of $\boldsymbol{\delta}({\bf L}^{\text M}_i )={\bf a}_i$ and are given by 
${\bf L}^{\text M}_i=({\mathcal I}-  {\mathcal R}^{-1}{\mathcal M}^{-1})^{-1}{\bf a}_i$ $(i=1,2)$.
The corresponding reciprocal vectors 
satisfying ${\bf G}^{\text M}_i\cdot {\bf L}^{\text M}_j = 2\pi\delta_{ij}$
are given by 
\begin{equation}
{\bf G}^{\text M}_i=({\mathcal I}-{\mathcal M}^{-1}{\mathcal R}){\bf b}_i \quad (i=1,2).	
\label{eq_g_def}
\end{equation}
We can immediately show that ${\bf G}^{\text M}_i\cdot {\bf C}=2\pi (n_i -n'_i)$ $(i=1,2)$ so that 
the moir\'e period is commensurate with the chiral vector ${\bf C}$ as it should. 
The periodic boundary condition for DWNT forces the two-dimensional wave space be quantized
into one-dimensional lines perpendicular to ${\bf C}$ with intervals of $2\pi/|{\bf C}|$.  

\begin{figure*}
	\begin{center}
		\leavevmode\includegraphics[width=0.95\hsize]{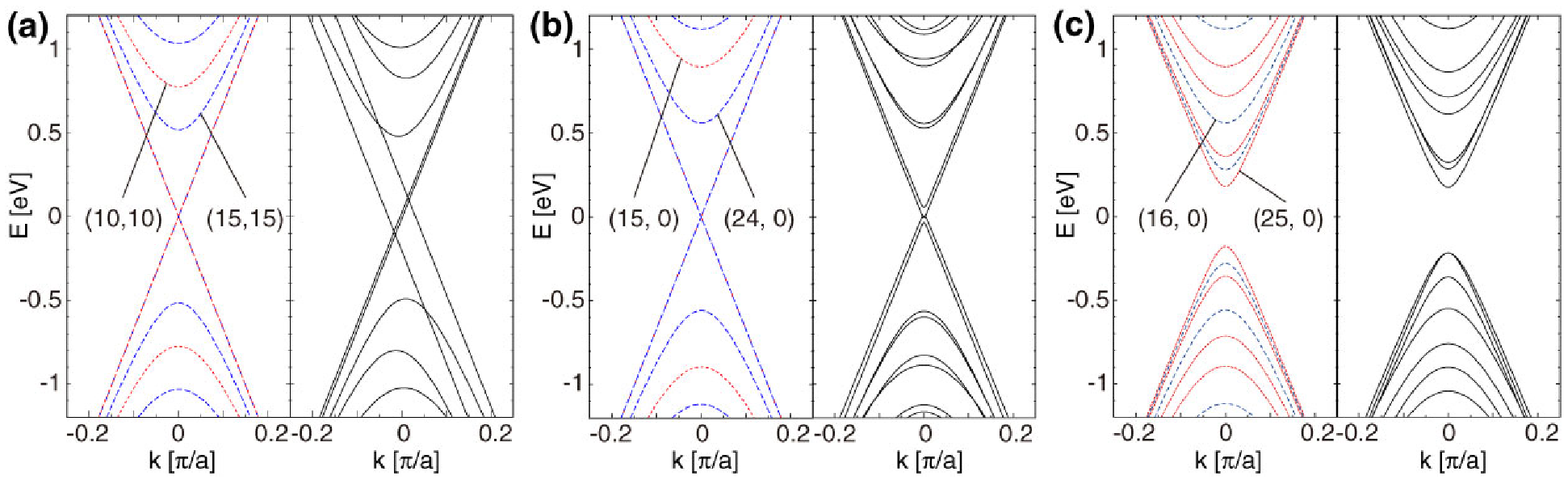}
	\end{center}
	\caption{
		Electronic energy band diagrams obtained by the atomic structure mapping and effective 
		theory are drawn for commensurate DWNTs. 
		(a) Energy bands of decoupled (10,10) and (15,15) SWNTs are drawn 
		in the left panel. Red (Blue) dotted lines are for inner (outer) tubes. In the right panel, the energy
		bands for coupled tube, i.e., (10,10)@(15,15) DWNT is drawn. 
		Same band diagrams for (15,0)@(24,0) and (16,0)@(25,0) DWNTs are drawn in (b)
		and (c), respectively. The former case is metallic and the latter semiconducting
		zigzag DWNT. In all band diagrams hereafter, the origin in x-axis for one-dimensional
		crystal momentum is taken at the $K$-point for metallic tubes and at the closest point to the 
		$K$-point for semiconducting tubes, respectively.
	}
	\label{fig_armchair_zigzag.ps}
\end{figure*}

\section{Effective Hamiltonian}

With the given conditions on the momentum spaces of the modified TBLG, 
now we construct the effective Hamiltonian for low energy electrons. 
The mapped lower layer for the outer tube has a distorted hexagonal Brillouin zone (BZ) while
the upper layer for the inner one has a usual BZ of graphene. 
Figures 2(a)-2(c) show the actual lattice structures and BZs
for each of DWNTs studied in the later sections. 
The low energy electrons can be described by effective Hamiltonians around the each corner of intralayer BZ:
${\bf K}_\xi=-\xi (2{\bf b}_1+{\bf b}_2)/3$ for the upper layer 
and $\tilde{\bf K}_\xi=-\xi(2\tilde{\bf b}_1+\tilde{\bf b}_2)/3={\mathcal M}^{-1}{\mathcal R}{\bf K}_\xi$ 
for the lower one where $\xi=\pm1$ denotes time-reversal partners.
Near the corners, the intralayer Hamiltonians for the upper and lower layer (layer 1 and 2 hereafter) 
can be written as
${\mathcal H}_1 ({\bf k})\simeq -\hbar v({\bf k}-{\bf K}_\xi)\cdot (\xi\sigma_x, \sigma_y)$ and 
${\mathcal H}_2 ({\bf k})\simeq-\hbar v[{\mathcal R}^{-1}{\mathcal M}({\bf k}-\tilde{\bf K}_\xi)]\cdot (\xi\sigma_x, \sigma_y)$, 
respectively
where ${\bf k}=(k_x,k_y)$ is the Bloch wave number for intralayer momentum space, $\sigma_x$ and $\sigma_y$
are the Pauli matrices acting on the two sublattices of upper ($A_1, B_1$) and lower ($A_2, B_2$) layer, 
and $v$ the electron velocity of graphene.
The low energy electrons of each layer interact through interlayer coupling such that 
the total Hamiltonian of the modified TBLG is written in the basis of $(A_1, B_1, A_2, B_2)$ as 
\begin{equation}
{\mathcal H}_\xi = 
\left(
\begin{array}{cc}
{\mathcal H}_1 ({\bf k})& U^\dagger \\
U & {\mathcal H}_2 ({\bf k})
\end{array}
\right),
\label{eq_H}
\end{equation}
where $U$ has interlayer coupling matrix elements expressed as 
$\langle {\bf k}', X'_{l'}|T|{\bf k}, X_l\rangle
$
where $|{\bf k}, X_l\rangle$ is an intralayer Bloch wave basis, 
$X$ and $X'$ are either
of $A$ or $B$,  $l$ and $l'$ are either of 1 or 2,
and 
$T$ is an interlayer coupling Hamiltonian.
In the following, we consider a situation where the moir\'{e} period is much greater than
the atomic scale, i.e., $|{\bf G}^\text{M}_i| \ll 2\pi/a$.
Then the interlayer matrix elements can be explicitly written in a quite simple form 
with three Fourier wave components of $1$, $e^{i\xi {\bf G}^\text{M}_1\cdot {\bf r}}$
and $e^{i\xi ({\bf G}^\text{M}_1 + {\bf G}^\text{M}_2)\cdot {\bf r}}$ as,
\begin{eqnarray}
&& U = 
\begin{pmatrix}
U_{A_2 A_1} & U_{A_2 B_1}
\\
U_{B_2 A_1} & U_{B_2 B_1}
\end{pmatrix}
=
u_0 (d)
\Biggl[
\begin{pmatrix}
1 & 1
\\
1 & 1
\end{pmatrix}
+
\nonumber\\
&&
\quad
\begin{pmatrix}
1 & \omega^{-\xi}
\\
\omega^{\xi} & 1
\end{pmatrix}
e^{i\xi\Vec{G}^{\rm M}_1\cdot\Vec{r}}
+
\begin{pmatrix}
1 & \omega^{\xi}
\\
\omega^{-\xi} & 1
\end{pmatrix}
e^{i\xi(\Vec{G}^{\rm M}_1+\Vec{G}^{\rm M}_2)\cdot\Vec{r}}
\Biggr],
\nonumber\\
\label{eq_U}
\end{eqnarray}
where $u_0$ is the coupling parameter depending on intertube distance of $d$,
and $\omega = \exp(2\pi i/3)$
(See derivations in Appendix \ref{app_interlayer_hamiltonian}).
We can infer that
the effect of interlayer coupling will be significant 
when the distance between the two $K$-points
of each layer, 
$\Delta {\bf K}_\xi \equiv \tilde{\bf K}_\xi -{\bf K}_\xi =\xi (2{\bf G}^\textrm{M}_1
+{\bf G}^\textrm{M}_2)/3$, is close to either of 
the three Fourier wave components 
0, $\xi {\bf G}^\text{M}_1$ or 
$\xi ({\bf G}^\text{M}_1 + {\bf G}^\text{M}_2)$.
This condition actually corresponds to the strong coupling case
referred in the next section.

We note that the effective Hamiltonian in Eq.\ (\ref{eq_H})
shares the essential features
with those describing other two-dimensional moir\'e crystals such as 
TBLG as well as graphene on hBN. 
The only difference arises from the definition of 
the moir\'{e} reciprocal vectors ${\bf G}^\text{M}_i$ [Eq.\ (\ref{eq_g_def})],
where ${\mathcal M}={\mathcal I}$ for TBLG~\cite{moon2},
while 
${\mathcal M}$ is an equibiaxial expansion 
(unlike the uniaxial one in the present case)
for graphene on hBN\cite{moon3}. 
In spite of such a apparent similarity, 
the DWNT is not 
a rolled-up version of two-dimensional moir\'{e} crystals, because 
there are two degrees of freedom,
${\mathcal M}$ and ${\mathcal R}$, 
depending on the choice of the inner and outer SWNTs,
giving a variation in the relative angles and magnitudes
of moire reciprocal vectors.
In TBLG, for example, 
${\bf G}^\text{M}_1$ and ${\bf G}^\text{M}_2$
are always 120$^\circ$ rotation of each other
due to the three-fold rotational symmetry,
and then $\Delta {\bf K}_\xi$ never agrees with
either of three Fourier wave components. 
In DWNT, as shown in the next section, 
a wider parameter space allows this condition to be met,
and also other distinct situations 
which are hardly realized in two-dimensional bilayer systems.

In Eq.\ (\ref{eq_U}), we neglect the dependence on the relative offset
between the atomic positions in the two nanotubes,
assuming that the two tubes share the same in-plane atomic position 
at some particular point.
In incommensurate DWNTs, the relative offset corresponds to 
a shift of the origin of the coordinate and does not change the electronic structure.
In commensurate DWNTs such as the zigzag-zigzag or armchair-armchair tubes, 
on the other hand, it should be noted that
the band structure generally does depend
on the offset, and it leads to a noticeable difference particularly in
small DWNTs.\cite{kwon1998electronic}

\begin{figure*}
	\begin{center}
		\leavevmode\includegraphics[width=0.8\hsize]{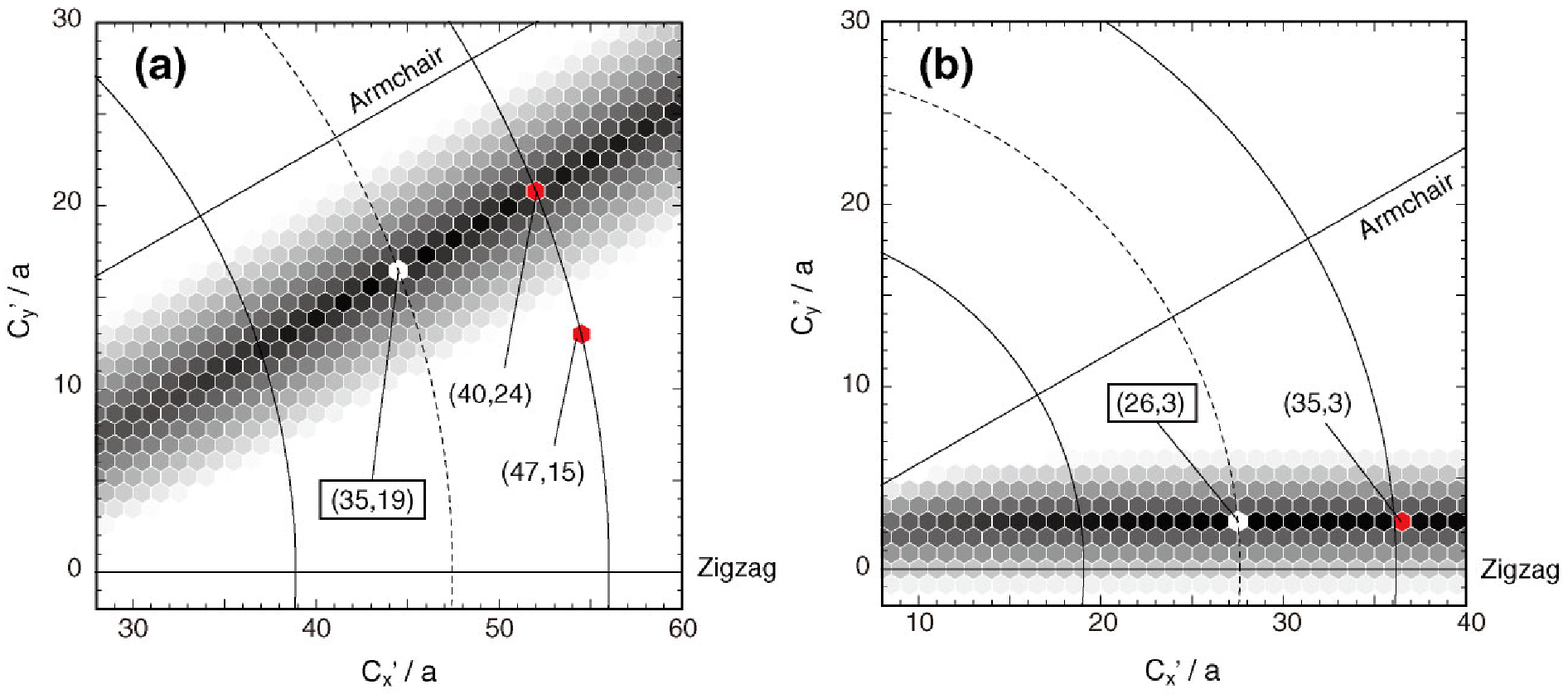}
	\end{center}
	\caption{
		(a) Two-dimensional distance map of $|\Delta {\bf K}_\xi - \xi{\bf G}_1^\text{M}|$
		by varying ${\bf C}'$ while ${\bf C}$ is fixed to (35,19) SWNT.
		Here the darker color indicates smaller distance and semicircle lines indicate 
		same radius of SWNTs.
		(b) Two-dimensional map of length of ${\bf G}^\text{M}_2$ 
		as a function of ${\bf C}'$ with ${\bf C}$ fixed to (26,3) SWNT, 
		where the darker color indicates smaller length.
	}
	\label{fig_strong_map.ps}
\end{figure*}

\section{Armchair and zigzag DWNTs}

By numerically solving eigenvalues of equation (\ref{eq_H})
under the quantization condition of 
${\bf k}\cdot {\bf C}=2\pi N$ ($N$ is integer),
we can obtain the energy-momentum 
relationship of electrons in DWNTs with and without commensurability. 
First, the well-known 
results for commensurate DWNTs are reproduced by using our method (Fig.\ 3). 
In the case of a DWNT having $(n,n)$ SWNT
inside $(m,m)$ one [hereafter $(n,n)$@$(m,m)$ DWNT], 
the calculated band structures from our continuum 
model agree well with previous results from 
{\it ab initio} methods~\cite{charlier_review} [Fig.\ 3(a)].
For a $(n,0)$@$(m,0)$ DWNT, 
the agreement between results from both methods are also very good [Figs. 3(b) and 3(c)]. 
In the former case, the low energy band structures deform greatly such 
that the two linear crossing bands push up and downward due to intertube
interactions while in the latter no significant deformation can be noted. 
This sharp contrast can be understood by checking the coupling 
condition considered before. 
In the former case, $\Delta {\bf K}_\xi$ exactly coincides with $\xi {\bf G}^\textrm{M}_1$
so that all combinations of $(n,n)$ SWNTs are always in the strong 
coupling condition. 
In the latter case, we have ${\bf G}^\textrm{M}_2=0$ 
and $\Delta{\bf K}_\xi=(2/3)\xi{\bf G}^\textrm{M}_1$, so that
$\Delta{\bf K}_\xi$ does not coincide
with either of 0, $\xi {\bf G}^\text{M}_1$ or 
$\xi ({\bf G}^\text{M}_1 + {\bf G}^\text{M}_2)$,
thus being. in the weak coupling condition.

\begin{figure*}
	\begin{center}
		\leavevmode\includegraphics[width=0.7\hsize]{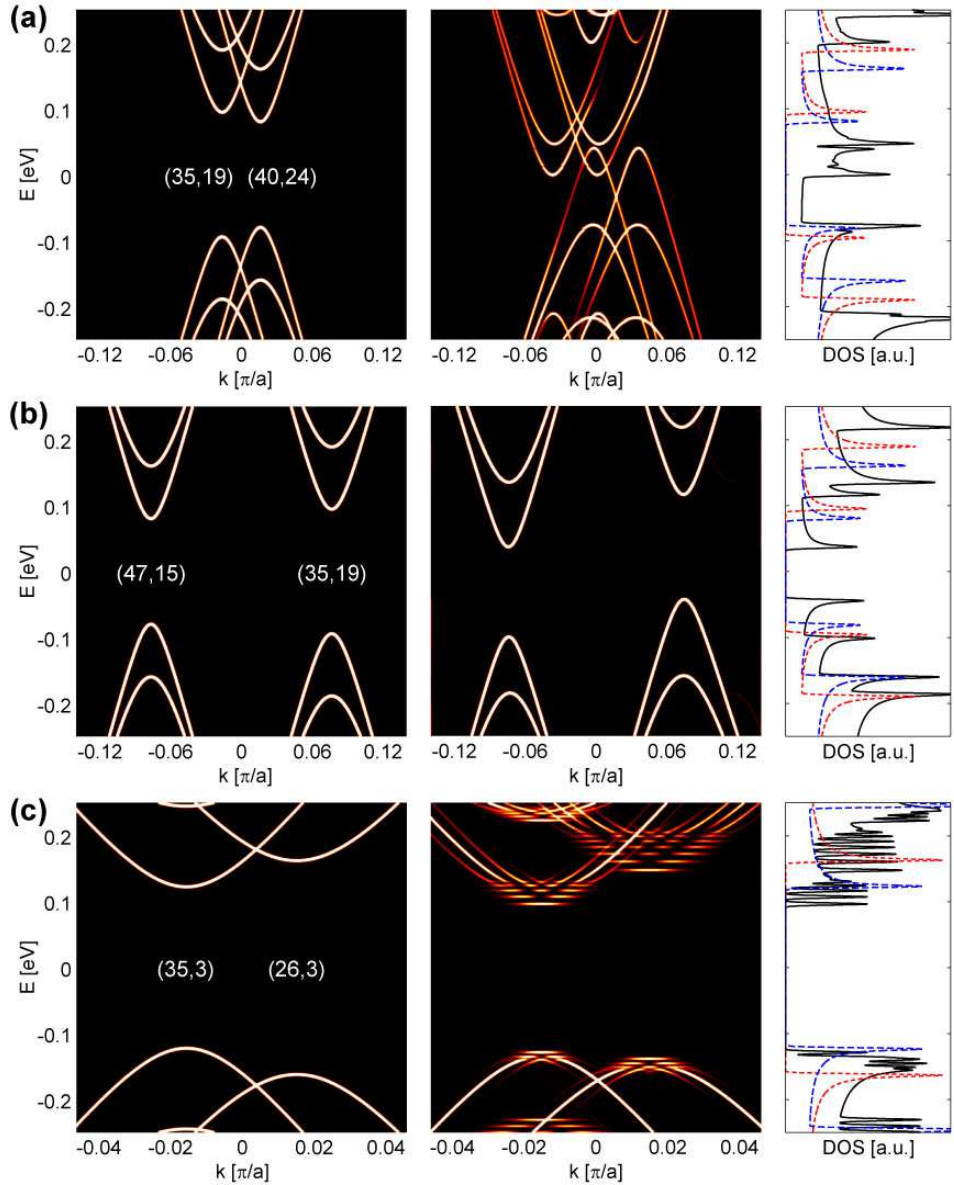}
	\end{center}
	\caption{
		Energy-momentum dispersion relation for incommensurate DWNTs are drawn with 
		the projected extend band plot method (See Appendix \ref{app_calculating_band}). 
		(a) Energy bands of decoupled (35,19) and (40,24) SWNTs are drawn 
		in the left panel. When the inter-tube coupling turns on, the dispersion of 
		(35,19)@(40,24) DWNT is shown in the middle panel. The corresponding density of 
		states (DOS) are drawn in the right panel. The dotted red (blue) lines are DOSs for 
		inner (outer) nanotubes and solid black line for the coupled DWNT.
		Same dispersion diagrams and DOS plots for (35,19)@(47,15) and (26,3)@(35,3) DWNTs
		are drawn in (b) and (c), respectively.
		The atomic lattice structures with distorted moir\'e lattices and corresponding BZs for all three
		cases are shown in Fig.\ 2. 
		For the both figures, a pair of panels in the same row 
		belongs to specific coupling conditions: from top to bottom rows,
		strong, weak, and localized coupling cases.
	}
	\label{fig_spectral_weight.ps}
\end{figure*}

\section{General Incommensurate DWNTs}

\subsection{Strong coupling condition}

The effective continuum model and the criteria for
the strong coupling work as well
for incommensurate and chiral DWNTs.
By measuring the distance between $\Delta{\bf K}_\xi$
and the three Fourier wavenumbers
with varying ${\bf C}$ and ${\bf C}'$,
we can find all possible combinations of SWNTs to make DWNTs 
with strong intertube couplings.
After some algebra, the criteria for the strong coupling is reduced to 
the simple conditions that
(i) ${\bf C}-{\bf C}'$ is parallel to the armchair direction,
(ii) ${\bf C}$ and ${\bf C}'$ are nearly parallel 
to each other
(See Appendix \ref{app_conditions} for the derivation).
For one example, here, we choose semiconducting (35,19) SWNT for the inner shell
and then search semiconducting outer SWNTs 
to show a strong or weak intertube coupling. 
Figure 4(a) shows the distance between $\Delta{\bf K}_\xi$
and $\xi {\bf G}^\text{M}_1$ 
as a function of ${\bf C}'$ with ${\bf C}$ fixed to (35,19),
where the darker color indicates smaller distance.
The strong coupling region actually
extends to the armchair direction
as expected from the criteria discussed before.
For the outer tube, we take (40,24) SWNT in the strong 
coupling condition, and (47,15) off from it, 
where the intertube distance is close to
the graphite's interlayer spacing in both cases. 
For (35,19)@(40,24) DWNT, 
the atomic structure of the corresponding modified TBLG
and its BZ are shown in Figs.\ 2(a) and 2(d), respectively. 
We see that ${\bf G}^\text{M}_1$ is indeed very close to
the displacement between two $K$-points in Fig.\ 2(d).

The calculated energy band structure is drawn with projected and
extended scheme in Fig.\ 5(a) 
(See Appendix \ref{app_calculating_band} for the calculation method).
Since the (35,19) tube has an energy band gap of 0.18 eV
and (40,24) of 0.15 eV (and the curvature effect
is too small to close the gap~\cite{okada}), 
one may expect that the DWNT
composed of the two tubes
will have an energy gap. 
However, the resulting band structure shows the characteristic
of metallic energy bands [Fig.\ 5(a)]. 
The lowest energy bands of decoupled nanotubes 
indeed mix together very strongly and the final 
low energy-momentum dispersions are quite different
from the original ones.
In the case of (35,19)@(47,15) DWNT [Fig.\ 2(b)], its energy-momentum
dispersion is nothing but a simple sum of the two tubes with a slight
energy shift [Fig.\ 5(b)] because this 
is off the strong coupling condition [Fig.\ 2(e)].
Corresponding density of states (DOS) for each case is displayed
in Figs. 5(a) and 5(b) showing a sharp contrast between the two coupling conditions, 
although these two DWNTs
have almost the same spectra in the absence of the intertube couping.
In the strong coupling regime, the interlayer Hamiltonian $U$
links the SWNT states at almost the same energy
and thus leads to an energy shift linear to $u_0$.
In general situations, on the other hand,
the two states connected by $u_0$
generally belong to different energies 
with a typical difference $\Delta E \sim \hbar v |{\bf G}^{\rm M}_i|$.
When $u_0 \ll \Delta E$,
the energy shift becomes the second order
as $\sim u_0^2/ \Delta E$, and this is the case in Fig.\ 5(b).


We can further obtain an insight from analytic expression
for energy gap of strongly coupled case.
The low energy spectrum of strongly coupled DWNT
is well approximated by the two Dirac cones separated by $\Delta{\bf K}_\xi$,
which are directly coupled by one of
three Fourier components,
0, $\xi {\bf G}^\text{M}_1$ or 
$\xi ({\bf G}^\text{M}_1 + {\bf G}^\text{M}_2)$.
Suppose that 
the lowest energy bands of decoupled inner and outer SWNTs 
with respect to each band center are expressed 
(here $\hbar v = 1$) as 
$E=\pm\left[m^2_\text{i}+k^2\right]^{1/2}$
and $E=\pm\left[m^2_\text{o}+ k^2\right]^{1/2}$
with energy gaps of
 $2|m_\text{i}|$ and $2|m_\text{o}|$, respectively.
When the two semiconducting SWNTs have similar diameters,
we can approximate $m_\text{i}, m_\text{o} \approx m$
and then the energy bands in the presence of the intertube coupling of $u_0$
are approximately written as four hyperbolas (See Appendix \ref{app_two-mode}).
If $\Delta{\bf K}_\xi\simeq\xi{\bf G}^\textrm{M}_1$, 
the four branches are given by,

 $$E(k) =-u_0 \pm \left[(m-m_D(u_0))^2+(k+k_D(u_0))^2 \right]^{1/2}$$
and 
$$E(k)=+u_0\pm\left[(m+m_D(u_0))^2+(k-k_D(u_0))^2\right]^{1/2},$$
where
$m_D(u_0)=u_0 \xi \cos(\phi+60^\circ)$,
$k_D(u_0)=u_0 \xi \sin(\phi+60^\circ)$,
and $\phi$ is the angle from $x$-axis to ${\bf C}$.
The energy gap is found to be $\Delta E = 2(|m|-u_0)$, 
and vanishes when $u_0 > |m|$.
From these expression, we find that the intertube interactions 
can indeed modify the semiconducting energy bands of bare SWNTs ~\cite{ywson}
into metallic ones in the strong coupling condition.
In Fig.\ \ref{fig_spectral_weight.ps},
we chose relatively large DWNTs (i.e. the energy gap is small)
such that $u_0 > |m|$, 
to actually demonstrate the gap closing.
Smaller DWNTs also have large band shifts in the strong
coupling condition, while they remain semiconducting
when the energy gaps of the constituent SWNTs are larger than $2u_0$.

\subsection{Localized insulating condition}

Besides the strong and weak coupling regimes, another 
classification is possible for the electronic structures of incommensurate DWNTs. 
In Figs.\ 2(c) and 2(f), we display a modified TBLG atomic structure and BZ for (26,3)@(35,3) DWNT,
and its energy-momentum dispersion and DOS
are shown in Fig.\ 5(c).
The two constituent SWNTs are both semiconducting, 
and their chiral vectors are almost 
parallel with $(n,0)$ nanotubes. 
Unlike previous two cases, we observe a number of flat bands 
both in conduction and valence energy bands, 
and the corresponding DOS also 
shows such characteristics [Fig.\ 5(c)]. 
The flat band occurs because electronic states at contiguous 
$k$-points on the same layer are hybridized by the
intertube coupling $U$.
Then an electron on each tube feels a periodic effective
potential with very long spatial period,
and the bound states appear at 
every single bottom of the effective potential.
The system is then viewed as a series of weakly connected quantum dots, and it offers a unique situation
in which identical quantum dots are arranged regularly at 
a precise period for a macroscopic length.

Since the matrix $U$ couples the different layers,
we need a second order process $U^\dagger G U$ or $U G U^\dagger$
($G$ is Green's function of decoupled SWNTs)
to connect the $k$-points on the same layer,
and such a process has the Fourier components of
$\pm{\bf G}^\text{M}_1$, $\pm{\bf G}^\text{M}_2$ and
$\pm({\bf G}^\text{M}_1+{\bf G}^\text{M}_2)$.
Therefore, the flat band localization condition requires
that either of 
${\bf G}^\text{M}_1$, ${\bf G}^\text{M}_2$ or ${\bf G}^\text{M}_1+{\bf G}^\text{M}_2$
is very small, but not exactly zero.
In the case of (26,3)@(35,3) DWNT,
$|{\bf G}^\text{M}_2|$ is merely about $0.014/a$, 
which corresponds to the spatial period about $450a\sim 110\,\mathrm{nm}$.
Similarly to the strong coupling case,
the criteria for the flat band is reduced to the simple conditions that
(i) ${\bf C}-{\bf C}'$ is parallel to the zigzag direction,
(ii) ${\bf C}$ and ${\bf C}'$ are nearly parallel 
(See Appendix \ref{app_conditions} for the derivation).
Figure 4(b) shows the length of ${\bf G}^\text{M}_2$ 
as a function of ${\bf C}'$ with the fixed ${\bf C}$ of $(26,3)$,
where the flat band region actually extends to the zigzag direction.
From the last consideration, we can conclude that DWNTs with two semiconducting
SWNTs can be classified into 
three categories, e.g., strong coupling near armchair-armchair DWNTs, 
localized insulating coupling near zigzag-zigzag ones and weak 
coupling cases otherwise.


\begin{figure*}
	\begin{center}
		\leavevmode\includegraphics[width=0.7\hsize]{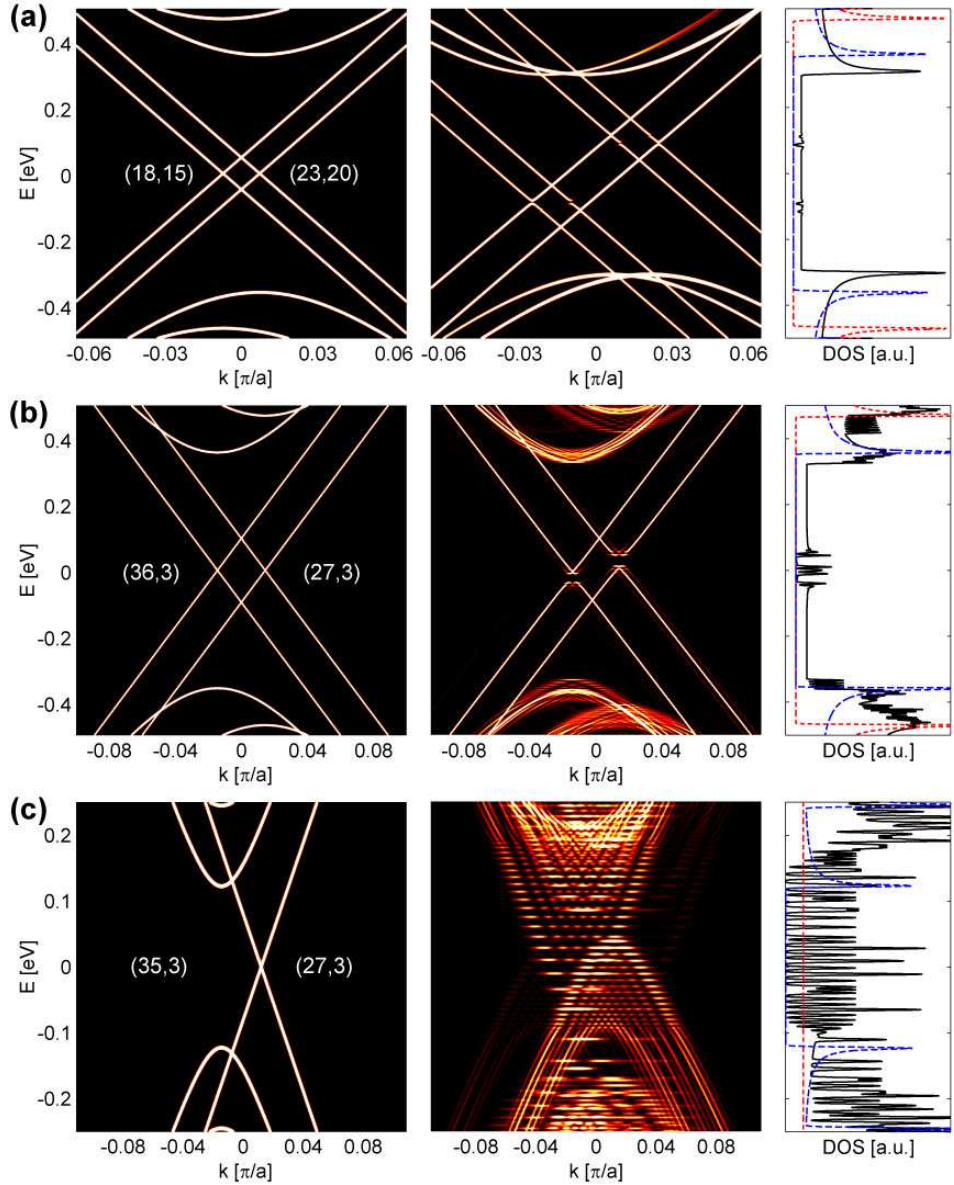}
	\end{center}
	\caption{
		(a)
		Energy bands of decoupled (18,15) and (23,20) SWNTs are drawn 
		in the left panel. When the inter-tube coupling is turn on, the dispersion of 
		(18,15)@(23,20) DWNT is shown in the middle panel. The corresponding density of 
		states (DOS) are drawn in the right panel. The dotted red (blue) lines are DOSs for 
		inner (outer) nanotubes and solid black line for the coupled DWNT.
		Same dispersion diagrams and DOS plots for (27,3)@(36,3) and (27,3)@(35,3) DWNTs
		are drawn in (b) and (c), respectively.
	}
	\label{fig_spectral_weight_metal}
\end{figure*}

\subsection{DWNTs including metallic SWNTs}

Our theory is not limited to semiconducting DWNTs.
Three coupling conditions still hold as well
when either or both of the two SWNTs are metallic.
Figure \ref{fig_spectral_weight_metal}(a) shows 
the spectrum of (18,15)@(23,20) DWNT in the strong-coupling condition,
where the low energy linear bands of decoupled metallic tubes
are repelled away without gap opening, similar to the armchair-armchair DWNT.
The insulating localized states are 
also possible for DWNTs composed of metallic SWNTs,
where the metallic behavior of the original SWNTs is completely lost
due to the formation of the bound states at the moir\'{e} potential extrema.
Figures \ref{fig_spectral_weight_metal}(b) and \ref{fig_spectral_weight_metal}(c) 
show the spectra for a DWNT consisting of two metallic SWNTs [(27,3)@(36,3)],
and for one consisting of a metallic SWNT and a semiconducting SWNT [(27,3)@(35,3)],  
respectively, both in the localized insulating condition.
We observe the formation of flat bands in both cases,
but in much greater energy range in (c) than in (b).
The significant difference comes from the different interlayer spacing $d$,
which gives the different interlayer coupling
$u_0(d)\approx 0.07$eV ($d\approx 0.351$nm) in the former
and 0.25 eV ($d\approx 0.312$nm) in the latter.
As the effective potential is the second order in $u_0$,
a change of the magnitude $u_0$ results in a significant difference
in the energy region where the flat bands are formed.
Actually the localized insulating condition strongly interferes with the 
condition for each SWNT to be metallic or semiconducting.
We can show that a metallic-metallic DWNT in the localized insulating condition 
appears only when $|\Vec{C}-\Vec{C}'|\approx 3ma$ with integer $m$,
and thus we have only a choice of $|\Vec{C}-\Vec{C}'|\approx 9a$ near the graphite interlayer spacing,
which is actually the case of Fig.\ \ref{fig_spectral_weight_metal}(b).


Finally, let us mention the relationship and differences between
one-dimensional and two-dimensional moir\'{e} crystals.
In the two-dimensional TBG, the strong interlayer coupling 
occurs only when the rotation angle is sufficiently small,
where the band dispersion of the low-lying levels are significantly suppressed\cite{trambly2010localization,morell2010flat,luican2011single,macdonald,moon1}. This situation corresponds to 
the strong coupling and localized insulating regimes 
in incommensurate DWNTs,
since the moire reciprocal lattice vectors
${\bf G}^{\rm M}_i$  and the K-point difference $\Delta{\bf K}_\xi$  
are all tiny and simultaneously satisfy the two conditions 
(for strong coupling and localized insulating) while not separately. 
On the other hand,
the uniqueness of one-dimensional moir\'{e} crystal is in that
the flexibility of moir\'{e} pattern due to more degrees of freedom
realizes the two different coupling conditions independently,
leading to a variety of situations 
which do not have explicit counterparts in two-dimensional systems.
For example, the strong coupling condition just requires 
$\Delta {\bf K}_\xi  \sim {\bf G}^{\rm M}_i$, but ${\bf G}^{\rm M}_i$  
is not necessarily small.
Therefore, as seen in Fig.\ \ref{fig_spectral_weight.ps}(a), 
we only have strong band repulsion due to the strong interlayer mixing, 
but the spectrum is not chopped into flat bands 
because ${\bf G}^{\rm M}_i$ is not small.
On the  other hand, the localized insulating regime actually requires that 
some ${\bf G}^{\rm M}_i$ is tiny, 
but not necessarily close to $\Delta {\bf K}_\xi$.
As a result, we have flat bands, but the hybridization between 1 and 2 is not strong as seen in Fig.\ \ref{fig_spectral_weight.ps}(c).

\section{Conclusions}

It is evident now that combination of SWNTs with almost the same
physical properties such as diameter and energy gap can end up with 
very different DWNTs depending on the interlayer moir\'{e} interference.
Therefore, all these criteria for incommensurate and chiral DWNTs
considered hitherto will dramatically influence 
their optical absorptions, photoluminescence, electric transport 
and Raman scattering that have been used for characterizing and understanding their
physical properties~\cite{saito,ando,charlier_review,shen,endo,shimamoto}.
Considering that the moir\'e pattern 
is present for almost all possible one-dimensional multishell tubular structures with 
several different atomic elements~\cite{tenne},
our current theoretical framework shall not be limited 
to multishell carbon nanotubes also.
Moreover, our study puts forth a new classification of nanotubes as the first example 
of one-dimensional moir\'e crystals
and paves a firm ground to utilize 
superb technological merits of DWNTs~\cite{shen,endo,shimamoto}.

{\it Note added}: After preparing this paper, we became aware of a recent 
paper~\cite{liu2014van} having a result that overlaps with a part of our theory.

\begin{acknowledgements}
M. K. was supported by JSPS Grant-in-Aid for Scientific
Research No. 24740193 and No. 25107005.
P. M. was supported by New York University Shanghai
Start-up Funds, and appreciate the support from
East China Normal University
for providing research facilities.
Y.-W.S. was supported by the NRF funded by the MSIP 
of Korean government (CASE, 2011-0031640 and QMMRC, No. R11-2008-053-01002-0).
Computations were supported by the CAC of KIAS.
\end{acknowledgements}

\appendix

\section{Interlayer Hamiltonian}
\label{app_interlayer_hamiltonian}

Here we derive the interlayer coupling matrix $U$
in the effective Hamiltonian of DWNT, Eq.\ (\ref{eq_H}) in the main text.
We assume that the moir\'{e} superlattice period 
is much larger than the lattice constant.
The local lattice structure is then approximately viewed as
a non-rotated bilayer graphene slided by
a displacement vector $\GVec{\delta}$, 
which slowly depends on the position
$\Vec{r}$ as
\begin{equation}
	\boldsymbol{\delta} ({\bf r})=({\mathcal I}-{\mathcal R}^{-1}{\mathcal M}^{-1}){\bf r}
	\label{eq_delta_r}
\end{equation} 
as argued in the main text.
Similarly to the two-dimensional moir\'{e} superlattice \cite{moon2,moon3}, 
the interlayer Hamiltonian of the DWNT is
obtained by replacing $\GVec{\delta}$ with $\GVec{\delta}(\Vec{r})$
in the Hamiltonian of non-rotated bilayer graphene
with a constant $\GVec{\delta}$.

Let us consider a non-rotated bilayer graphene 
with a constant in-plane displacement $\GVec{\delta}$
and interlayer spacing $d$.
We define $\Vec{a}_1$ and $\Vec{a}_2$ as the lattice vectors of graphene,
$\Vec{b}_1$ and $\Vec{b}_2$ as the corresponding reciprocal lattice vectors.
We model the system with the tight-binding model for $p_z$ atomic orbitals.
The Hamiltonian is written as
\begin{eqnarray}
	H = -\sum_{\langle i,j\rangle}
	t(\Vec{R}_i - \Vec{R}_j)
	|\Vec{R}_i\rangle\langle\Vec{R}_j| + {\rm H.c.},
	\label{eq_Hamiltonian_TBG}
\end{eqnarray}
where $\Vec{R}_i$ and $|\Vec{R}_i\rangle$ 
represent the lattice point and the atomic state at site $i$, respectively,
and $t(\Vec{R}_i - \Vec{R}_j)$ is
the transfer integral between the sites $i$ and $j$. 
We adopt a Slater-Koster parametrization \cite{slater1954simplified}
\begin{eqnarray}
	&& -t(\Vec{R}) = 
	V_{pp\pi}\left[1-\left(\frac{\Vec{R}\cdot\Vec{e}_z}{d}\right)^2\right]
	+ V_{pp\sigma}\left(\frac{\Vec{R}\cdot\Vec{e}_z}{d}\right)^2,
	\nonumber \\
	&& V_{pp\pi} =  V_{pp\pi}^0 e^{- (R-a_0)/r_0},
	\quad V_{pp\sigma} =  V_{pp\sigma}^0  e^{- (R-d_0)/r_0},
	\nonumber\\ 
	\label{eq_transfer_integral}
\end{eqnarray}
where
$\Vec{e}_z$ is the unit vector
perpendicular to the graphene plane,
$a_0 = a/\sqrt{3} \approx 0.142\,\mathrm{nm}$ is the distance of
neighboring $A$ and $B$ sites on monolayer,
and $d_0 \approx 0.335\,\mathrm{nm}$
is the interlayer spacing if bulk graphites.
Other parameters are typically
$V_{pp\pi}^0 \approx -2.7\,\mathrm{eV}$,
$ V_{pp\sigma}^0 \approx 0.48\,\mathrm{eV}$
and $r_0 \approx 0.045\,\mathrm{nm}$. \cite{moon2}


We define the Bloch wave basis of a single layer as
\begin{eqnarray}
	&& |\Vec{k},X_l\rangle = 
	\frac{1}{\sqrt{N}}\sum_{\Vec{R}_{X_l}} e^{i\Vec{k}\cdot\Vec{R}_{X_l}}
	|\Vec{R}_{X_l}\rangle,
	\label{eq_bloch_base}
\end{eqnarray}\
where $X = A,B$ is the sublattice index, $l= 1, 2$ is the layer index,
and $N$ is the number of monolayer's unit cell in the whole system.
The interlayer matrix element is then written as
\begin{eqnarray}
	&& U_{A_2A_1}(\Vec{k},\GVec{\delta}) 
	\equiv \langle \Vec{k},A_2| H |\Vec{k},A_1\rangle
	= u(\Vec{k},\GVec{\delta}),
	\nonumber\\
	&& U_{B_2B_1}(\Vec{k},\GVec{\delta}) 
	\equiv \langle \Vec{k},B_2| H |\Vec{k},B_1\rangle
	= u(\Vec{k},\GVec{\delta}),
	\nonumber\\
	&& U_{B_2A_1}(\Vec{k},\GVec{\delta}) 
	\equiv \langle \Vec{k},B_2| H |\Vec{k},A_1\rangle
	= u(\Vec{k},\GVec{\delta} - \GVec{\tau}_1),
	\nonumber\\
	&& U_{A_2B_1}(\Vec{k},\GVec{\delta}) 
	\equiv \langle \Vec{k},A_2| H |\Vec{k},B_1\rangle
	= u(\Vec{k},\GVec{\delta} + \GVec{\tau}_1),
	\label{eq_interlayer_U}
\end{eqnarray}
where
\begin{eqnarray}
	u(\Vec{k},\GVec{\delta}) = 
	\sum_{n_1,n_2}
	- t(n_1 \Vec{a}_1 + n_2 \Vec{a}_2 + d\Vec{e}_z + \GVec{\delta})
	\nonumber\\
	\hspace{20mm}
	\times
	\exp\left[-i\Vec{k}\cdot(n_1 \Vec{a}_1 + n_2 \Vec{a}_2 + \GVec{\delta})
	\right].
\end{eqnarray}
Here $\GVec{\tau}_1 = (-\Vec{a}_1+2\Vec{a}_2)/3$ is a vector
connecting the nearest $A$ and $B$ sublattices,
and $\Vec{e}_z$ is the unit vector perpendicular to the graphene plane.

Since the function $u(\Vec{k},\GVec{\delta})$ is 
periodic in $\GVec{\delta}$ with periods $\Vec{a}_1$ and $\Vec{a}_2$,
it is Fourier transformed as,
\begin{eqnarray}
	u(\Vec{k},\GVec{\delta}) = 
	-\sum_{m_1,m_2} \tilde{t}(m_1\Vec{b}_1+m_2\Vec{b}_2+\Vec{k})
	\nonumber\\
	\hspace{20mm}
	\times
	\exp[
	i(m_1\Vec{b}_1+m_2\Vec{b}_2)\cdot \GVec{\delta}
	],
	\label{eq_ukq_2}
\end{eqnarray}
where $\tilde{t}(\Vec{q})$ is the in-plane Fourier transform
of $t(\Vec{R})$ defined by
\begin{eqnarray}
	\tilde{t}(\Vec{q}) = 
	\frac{1}{S} \int t(\Vec{R}+ d\Vec{e}_z) 
	e^{-i \Vec{q}\cdot \Vec{R}} d\Vec{R},
	\label{eq_tilde_t}
\end{eqnarray}
with $S = |\Vec{a}_1\times\Vec{a}_2|$, and the integral in $\Vec{R}$
is taken over an infinite two-dimensional space.
In the present tight-binding model,
$t(\Vec{R})$ exponentially decays in $R \, \gsim \, r_0$, so that
the Fourier transform $\tilde{t}(\Vec{q})$ decays in $q \, \gsim \, 1/r_0$.
In Eq.\ (\ref{eq_ukq_2}), therefore, we only need to take
a few Fourier components within 
$|m_1\Vec{b}_1+m_2\Vec{b}_2+\Vec{k}| \,\lsim\, O(1/r_0)$.

\begin{figure}
	\begin{center}
		\leavevmode\includegraphics[width=0.9\hsize]{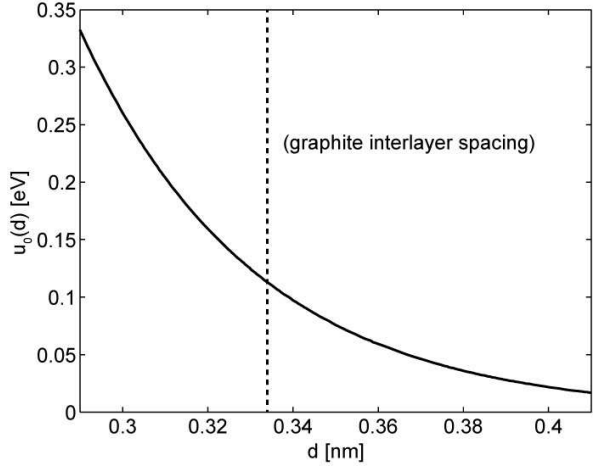}
	\end{center}
	\caption{
		Dependence of $u_0$ on
		interlayer spacing $d$.
	}
	\label{fig_u0}
\end{figure}

In the following 
we only consider the electronic states near $\Vec{K}_\xi$ point,
and then we can approximate $u(\Vec{k},\GVec{\delta})$
with $u(\Vec{K}_\xi,\GVec{\delta})$.
Eq.\ (\ref{eq_ukq_2}) then becomes
\begin{eqnarray}
	u(\Vec{K}_\xi,\GVec{\delta})  \approx 
	u_0 \left[
	1 + e^{i \xi \Vec{b}_1 \cdot \mbox{\boldmath \scriptsize $\delta$}}
	+ e^{i \xi (\Vec{b}_1 + \Vec{b}_2)\cdot  \mbox{\boldmath \scriptsize $\delta$}}
	\right],
\end{eqnarray}
with
\begin{eqnarray}
	u_0 = \tilde{t}(\Vec{K}_\xi).
\end{eqnarray}
Note that $u_0$ depends on interlayer spacing $d$
through $\tilde{t}(\Vec{q})$ in Eq.\ (\ref{eq_tilde_t}).
In the present choice of the tight-binding parameters
we have $u_0 = 0.11$eV 
at the graphite interlayer spacing, $d = 0.334\,\mathrm{nm}$. 
The second largest Fourier component
is $\tilde{t}(2\Vec{K}_\xi) \approx 0.0016$eV,
and is safely neglected.
Unlike the graphite system,
DWNTs can have wide range
of $d$ between $0.29\, \mathrm{nm}$ and $0.41\, \mathrm{nm}$
\cite{villalpando2010tunable,pfeiffer2006tube,villalpando2008raman,ren2002morphology}.
In this range,
$u_0$ also varies widely
from $0.33\, \mathrm{eV}$ to $0.017\, \mathrm{eV}$,
as we plot in Fig.\ \ref{fig_u0}.
By replacing $\GVec{\delta}$ with $\GVec{\delta}(\Vec{r})$
in Eq.\ (\ref{eq_delta_r}), we obtain
the interlayer Hamiltonian of the DWNT, Eq.\ (\ref{eq_U}).
Here we used the relation
$
\Vec{b}_i \cdot \GVec{\delta}(\Vec{r})
= \Vec{G}^{\rm M}_i\cdot\Vec{r}.
$

\section{Conditions for strong coupling case and flat band case}
\label{app_conditions}

We derive the condition for the two chiral vectors
$\Vec{C}$ and $\Vec{C}'$ to give the strong coupling case
and the flat band case.
The strong interlayer coupling occurs when
$\Delta \Vec{K}_\xi = \xi(2\Vec{G}_1^M + \Vec{G}_2^M)/3$
is close to either of 
$0$, $\xi\Vec{G}_1^M$ or $\xi(\Vec{G}_1^M+\Vec{G}_2^M)$
(see the main text).
The condition is written as
\begin{eqnarray}
	\begin{cases}
		2 \Vec{G}_1^M + \Vec{G}_2^M \approx 0 & {\rm or}
		\\
		\Vec{G}_1^M - \Vec{G}_2^M \approx 0  & {\rm or}
		\\
		\Vec{G}_1^M + 2\Vec{G}_2^M \approx 0. 
	\end{cases}
\end{eqnarray}
Using $\Vec{G}_i^M = ({\mathcal I}-{\mathcal M}^{-1}{\mathcal R})\Vec{b}_i$, 
this is rewritten as
\begin{eqnarray}
	\begin{cases}
		({\mathcal I}-{\mathcal M}^{-1}{\mathcal R})(2 \Vec{b}_1 + \Vec{b}_2) \approx 0 & {\rm or}
		\\
		({\mathcal I}-{\mathcal M}^{-1}{\mathcal R})(\Vec{b}_1 - \Vec{b}_2) \approx 0  & {\rm or}
		\\
		({\mathcal I}-{\mathcal M}^{-1}{\mathcal R})(\Vec{b}_1 + 2\Vec{b}_2) \approx 0.
	\end{cases}
	\label{eq_cond_for_strong}
\end{eqnarray}
Since the vectors 
$2 \Vec{b}_1 + \Vec{b}_2$, 
$\Vec{b}_1 - \Vec{b}_2$, 
and $\Vec{b}_1 + 2\Vec{b}_2$ 
are parallel to zigzag direction
(i.e., $0$, $2\pi/3$, $-2\pi/3$ from the $x$-axis),
the condition Eq.\ (\ref{eq_cond_for_strong}) is simplified to
\begin{eqnarray}
	({\mathcal I}-{\mathcal M}^{-1}{\mathcal R})\Vec{x} \approx 0
\end{eqnarray}
for $\Vec{x}$ parallel to a zigzag direction.

\begin{figure}
	\begin{center}
		\leavevmode\includegraphics[width=0.95\hsize]{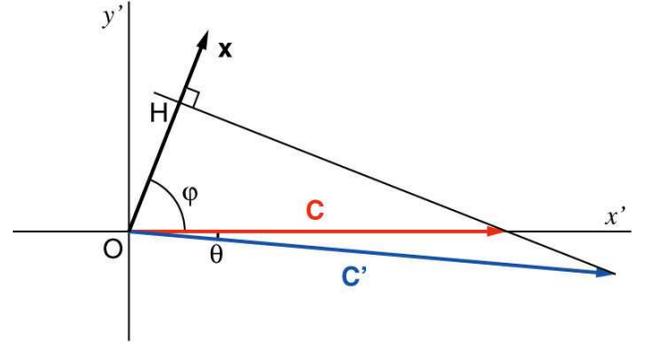}
	\end{center}
	\caption{Geometry to explain the strong coupling condition
		(see the text).
	}
	\label{fig_strong_cond}
\end{figure}

Let us consider the amplitude of $({\mathcal I}-{\mathcal M}^{-1}{\mathcal R})\Vec{x}$
as a function of $\Vec{x}$.
We introduce a rotated coordinate system $(x',y')$
with $x'$-axis set to parallel to $\Vec{C}$.
Then we can write
\begin{eqnarray}
	M = \begin{pmatrix}
		C/C' & 0 \\ 0 &  1
	\end{pmatrix},
	\quad
	R = \begin{pmatrix}
		\cos\theta & -\sin\theta \\ \sin\theta & \cos\theta  
	\end{pmatrix}.
\end{eqnarray}
For $\Vec{x}=(\cos\varphi,\sin\varphi)$,
we have
\begin{eqnarray}
	&& |({\mathcal I}-{\mathcal M}^{-1}{\mathcal R})\Vec{x}|^2 =
	\left(\frac{C'}{C}\cos(\varphi +\theta)-\cos\varphi\right)^2
	\nonumber
	\\
	&&\hspace{30mm}
	+(\sin(\varphi +\theta)-\sin\varphi)^2. 
	\label{eq_1-mr}
\end{eqnarray}
The first term in the right hand side vanishes when 
$\Vec{x}$ is perpendicular to $\Vec{C}-\Vec{C'}$.
This is geometrically explained in Fig.\ \ref{fig_strong_cond},
where we actually see 
${\rm OH}=C'\cos(\varphi +\theta) = C\cos\varphi$
when $\Vec{x} \perp \Vec{C}-\Vec{C'}$.
The second term  becomes small when $\theta$ is small,
i.e. $\Vec{C}$ and $\Vec{C}'$ are nearly parallel.
Therefore, we have strong intertube coupling when
(i) $\Vec{C}-\Vec{C'}$ is parallel to the armchair direction
(i.e., perpendicular to the zigzag direction), and 
(ii) $\Vec{C}$ and $\Vec{C'}$ are nearly parallel.

On the other hand, the flat band case
takes place when
either of 
$\Vec{G}_1^M$, $\Vec{G}_2^M$ or $\Vec{G}_1^M+\Vec{G}_2^M$
is very close to zero, but not exactly zero.
In a similar manner, the condition is rewritten as
\begin{eqnarray}
	\begin{cases}
		({\mathcal I}-{\mathcal M}^{-1}{\mathcal R}) \Vec{b}_1 \approx 0 & {\rm or}
		\\
		({\mathcal I}-{\mathcal M}^{-1}{\mathcal R}) \Vec{b}_2 \approx 0  & {\rm or}
		\\
		({\mathcal I}-{\mathcal M}^{-1}{\mathcal R})(\Vec{b}_1 + \Vec{b}_2) \approx 0.
	\end{cases}
	\label{eq_cond_for_flat}
\end{eqnarray}
Since the vectors 
$\Vec{b}_1$, 
$\Vec{b}_2$, 
and $\Vec{b}_1 + \Vec{b}_2$ 
are now parallel to the armchair direction,
the flat band condition is obtained by replacing
"zigzag" and "armchair" in the previous argument for the strong coupling condition.
Therefore, we have a flat band DWNT when
(i) $\Vec{C}-\Vec{C'}$ is parallel to the zigzag direction, and 
(ii) $\Vec{C}$ and $\Vec{C'}$ are nearly parallel.

\section{Calculating band structures of chiral DWNTs}
\label{app_calculating_band}

\begin{figure}
	\begin{center}
		\leavevmode\includegraphics[width=0.9\hsize]{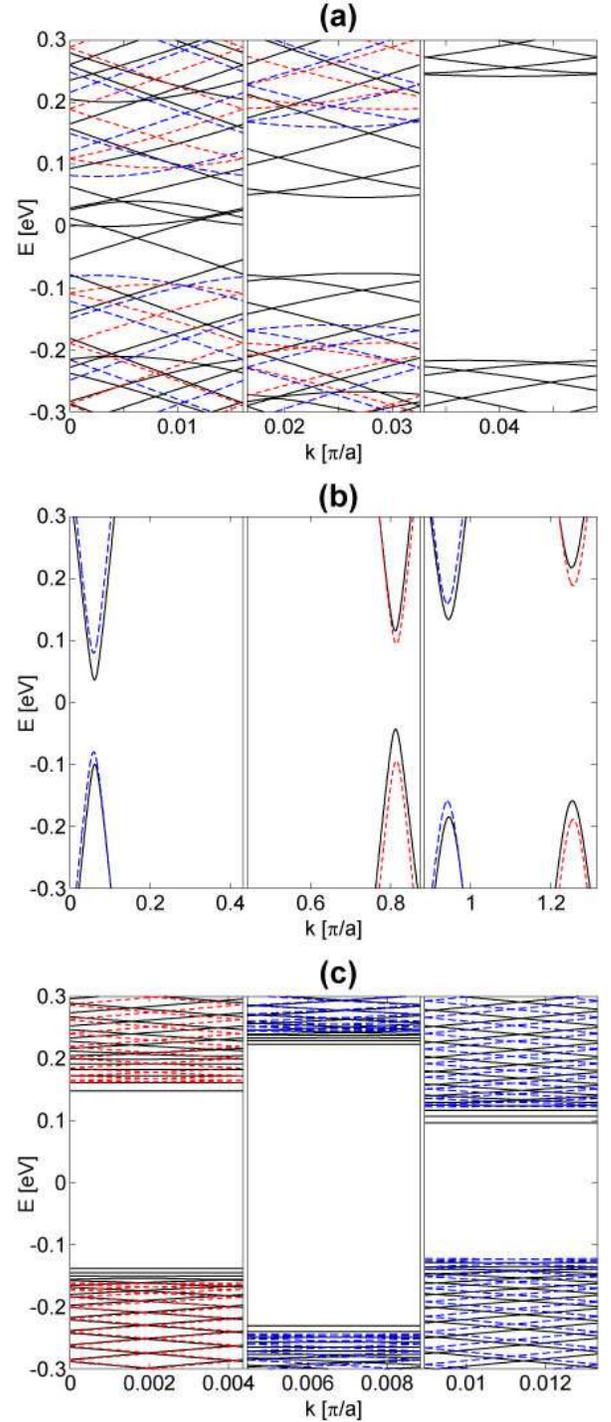}
	\end{center}
	\caption{
		Band structures near ${\bf K}_+$ valley of DWNTs with
	    (a) (35,19)@(40,24) (strong coupling), 
		(b) (35,19)@(47,15) (weak coupling),
		and (c) (26,3)@(35,3) (flat band).
		The solid (black) lines represent the energy bands of
		the DWNT in the presence interlayer coupling,
		while the dotted red (blue)
		lines are those of inner (outer) SWNTs.
		We do not show the subgroups
		which contain no energy bands
		in the given range.
	}
	\label{fig_band_structures}
\end{figure}

We present the details of the band calculation 
for DWNT with the effective continuum theory.
Every eigenstate is labeled by the Bloch wave number $\Vec{k}$
defined on the cutting lines 
${\bf k}\cdot {\bf C}=2\pi N$ ($N$ is integer)
inside the two-dimensional Brillouin zone
spanned by $\Vec{G}_1^{\rm M}$ and $\Vec{G}_2^{\rm M}$.
Since ${\bf G}^{\text M}_i\cdot {\bf C}=2\pi (n_i -n'_i)$,
the cutting lines are categorized into
$n_r$ different subgroups where $n_r={\rm GCD}(n_1-n'_1,n_2-n'_2)$.
To obtain the energy spectrum, we take the $k$-points of
$\Vec{q}=\Vec{k} +m_1\Vec{G}_1^{\rm M}+m_2\Vec{G}_2^{\rm M}$
($m_1,m_2$: integers) in the region 
$|\Vec{q}-(\Vec{K}_\xi+\tilde{\Vec{K}}_\xi)/2| < k_{\rm max}$
with a sufficiently large wave-cutoff $k_{\rm max}$, 
and numerically diagonalize the Hamiltonian within the limited wave space. 
Figure \ref{fig_band_structures} shows the band structures of $\xi=+$ valley
calculated for DWNTs studied in the main text;
(a) (35,19)@(40,24),
(b) (35,19)@(47,15), and (c) (26,3)@(35,3).
Here the energy bands are separately plotted for each of $n_r$ subgroups,
while we omitted the subgroups which contain no energy bands in the given range.
The solid curves represent the energy bands of
the DWNT with the interlayer coupling, and the dotted and dashed
curves are those of independent SWNTs without coupling.
We actually see the band gap closing in the strong coupling case
[Fig.\ \ref{fig_band_structures}(a)]
and the flat low-energy bands in the flat band case
[Fig.\ \ref{fig_band_structures}(c)]
as argued in the main text.

In Figs.\ \ref{fig_spectral_weight.ps} and \ref{fig_spectral_weight_metal} 
in the main text, we presented the spectral function
in the extended zone scheme
instead of the complex band structure folded into the first Brillouin zone.
This is defined as
\begin{equation}
	A(\Vec{k},\varepsilon)	
	= \sum_\alpha \sum_{X,l} 
	|\langle \alpha | \Vec{k}, X_l\rangle|^2 \delta(\varepsilon-\varepsilon_\alpha),
\end{equation}
where $|\alpha\rangle$ and $\varepsilon_\alpha$
are the eigenstate and the eigenenergy, respectively,
$X = A,B$ is the sublattice index, $l= 1, 2$ is the layer index, and
$|\Vec{k}, X_l\rangle$ is the plane wave basis defined by Eq.\ (\ref{eq_bloch_base}).
The spectral function is defined on the cutting lines
${\bf k}\cdot {\bf C}=2\pi N$ on the infinite two-dimensional reciprocal space,
and not limited to the reduced Brillouin zone.
Figures \ref{fig_spectral_weight.ps} and \ref{fig_spectral_weight_metal} 
are obtained by
taking summation of the spectral functions over different cutting lines
near a single $K_\xi$ point, and projecting it on a single $k$-axis.

\section{Two-mode approximation in strong interlayer coupling condition}
\label{app_two-mode}

\begin{figure}
	\begin{center}
		\leavevmode\includegraphics[width=0.65\hsize]{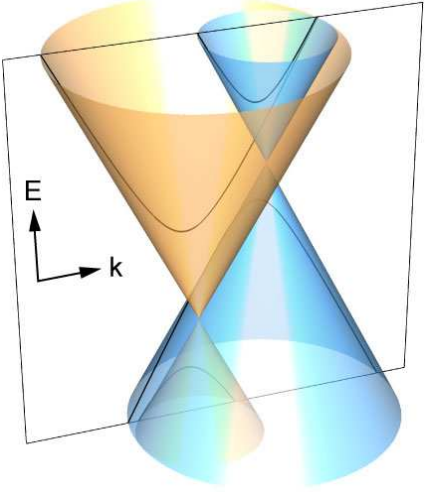}
	\end{center}
	\caption{
		Lowest energy bands of DWNT in strong coupling condition.
		The surface plot shows the energy dispersion
		of the modified TBLG, Eq.\ (\ref{eq_twomode_soln_2D}).
		The black lines show the one-dimensional dispersion
		Eq.\ (\ref{eq_E1D_two_modes_approx})
		along the quantization line of semiconducting DWNT (see text).
	}
	\label{fig_twomode_3D}
\end{figure}

\begin{figure*}
	\begin{center}
		\leavevmode\includegraphics[width=0.7\hsize]{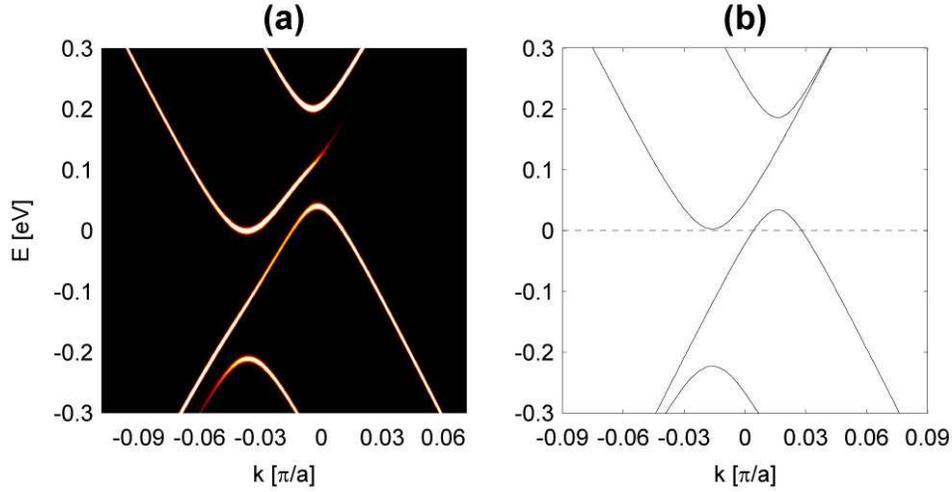}
	\end{center}
	\caption{
		(a) Numerically calculated band dispersions
		in the extended zone scheme for (35,19)@(40,24).
		(b) Band dispersions of the same DWNT
		calculated by the analytic expression
		[Eq.\ (\ref{eq_E1D_two_modes_approx})].
	}
	\label{fig_twomode_1D}
\end{figure*}

Here we derive an approximate analytic expression
of the low energy spectrum of DWNTs
in the strong coupling condition.
We consider the strong coupling case of $\Delta\Vec{K}_\xi = \xi\Vec{G}_1^{\rm M}$,
and apply the two-mode approximation
for the two Dirac cones of layer 1 and 2 
which are directly coupled by one of the three Fourier components,
$\xi {\bf G}^\text{M}_1$,
in the interlayer Hamiltonian.
The effective Hamiltonian is written as
\begin{eqnarray}
	\mathcal{H}_\mathrm{low} = 
	\begin{pmatrix}
		\mathcal{H}_1(\Vec{k})
		& U^\dagger
		\\
		U
		& \mathcal{H}_2(\Vec{k})
	\end{pmatrix},
	\label{eq_twomode_full}
\end{eqnarray}
where
\begin{eqnarray}
	&& \mathcal{H}_1(\Vec{k}) \simeq - \hbar v(\Vec{k}-\Vec{K}_\xi)\cdot (\xi\sigma_x, \sigma_y),
	\nonumber\\
	&& \mathcal{H}_2(\Vec{k}) \simeq - \hbar v[\mathcal{R}^{-1}\mathcal{M}(\Vec{k}-{\Vec{K}}_\xi-\Delta{\Vec{K}}_\xi)]\cdot (\xi\sigma_x, \sigma_y),
	\nonumber\\
	&& U = u_0
	\begin{pmatrix}
		1
		& \omega^{-\xi}
		\\
		\omega^{\xi}
		& 1
	\end{pmatrix}e^{i\xi\Vec{G}^{\rm M}_1\cdot\Vec{r}}.
\end{eqnarray}
The two Dirac cones are separated by  $\Delta\Vec{K}_\xi$,
and they are exactly merged by the Fourier component $e^{i\xi\Vec{G}^{\rm M}_1\cdot\Vec{r}}$
since $\Delta\Vec{K}_\xi = \xi\Vec{G}_1^{\rm M}$.
By applying a unitary transformation 
$\mathcal{H}'_\mathrm{low} = V^{\dagger}\mathcal{H}_\mathrm{low} V$
with 
$V={\rm diag}(1,1,e^{i\xi\Vec{G}^{\rm M}_1\cdot\Vec{r}},e^{i\xi\Vec{G}^{\rm M}_1\cdot\Vec{r}})$,
Eq.\ (\ref{eq_twomode_full}) is simplified to
\begin{eqnarray}
	\mathcal{H}'_\mathrm{low} =
	\begin{pmatrix}
		\mathcal{H}'(\Vec{k})
		& U'^\dagger
		\\
		U'
		& \mathcal{H}'(\Vec{k})
	\end{pmatrix},
	\label{eq_twomode_simple}
\end{eqnarray}
with
\begin{align}
	& \mathcal{H}'(\Vec{k}) = - \hbar v\Vec{k}\cdot (\xi\sigma_x, \sigma_y),
	\nonumber\\
	& U' = u_0
	\begin{pmatrix}
		1
		& \omega^{-\xi}
		\\
		\omega^{\xi}
		& 1
	\end{pmatrix},
\end{align}
where the wave number $\Vec{k}$
is measured relative to $\Vec{K}_\xi$.
and we use the approximation $\mathcal{R}^{-1}\mathcal{M}\Vec{k}\approx \Vec{k}$
assuming that $\mathcal{R}^{-1}\mathcal{M}$ is close to the identity matrix,
i.e., $\Vec{C}$ and $\Vec{C}'$ sufficiently close to each other.
The above equation gives the energy dispersions
of two shifted Dirac cones
\begin{eqnarray}
	&& E^\pm_1(\Vec{k}) =  -u_0 \pm \hbar v |\Vec{k}-\Vec{k}_0|,
	\nonumber\\
	&& E^\pm_2(\Vec{k}) = u_0 \pm \hbar v |\Vec{k}+\Vec{k}_0|,
	\label{eq_twomode_soln_2D}
\end{eqnarray}
where $\Vec{k} = (k_x, k_y)$ and
\begin{eqnarray}
	\Vec{k}_0
	\equiv  \frac{u_0\xi}{\hbar v}
	\begin{pmatrix}
		\cos(-60^\circ)
		\\
		\sin(-60^\circ)
	\end{pmatrix}.
\end{eqnarray}
The surface plot in Fig.\ \ref{fig_twomode_3D}
shows the dispersion Eq.\ (\ref{eq_twomode_soln_2D}),
where we see that the two shifted Dirac cones
touch on a single line $E = - \hbar v {\bf k}\cdot{\bf k}_0/|{\bf k}_0|$.

The lowest energy bands of DWNTs
along the quantization line
closest to $\Vec{K}_\xi$ are given as
\begin{eqnarray}
	\Vec{k} = 
	k
	\begin{pmatrix}
		-\sin\phi
		\\
		\cos\phi
	\end{pmatrix}
	+
	m
	\begin{pmatrix}
		\cos\phi
		\\
		\sin\phi
	\end{pmatrix},
\end{eqnarray}
where
$\phi$ is the angle from $x$-axis to $\Vec{C}$,
$k$ is one-dimensional wave number
along the tube axis, 
$m = 2\pi\nu\xi/(3C)$ and  
$\nu = 2n_1+n_2$ (in modulo of 3) is either of 0, 1 or $-1$.
This gives four branches of one-dimensional
energy bands
\begin{eqnarray}
	&& E^\pm_1(k) = -u_0 \pm \hbar v \sqrt{(m-m_D(u_0))^2
		+(k+k_D(u_0))^2},
	\nonumber\\
	&& E^\pm_2(k) = u_0 \pm \hbar v \sqrt{(m+m_D(u_0))^2
		+(k-k_D(u_0))^2},
	\nonumber\\
	\label{eq_E1D_two_modes_approx}
\end{eqnarray}
where
\begin{eqnarray}
	&& m_D(u_0) \equiv \frac{u_0\xi}{\hbar v} \cos(\phi+60^\circ),
	\nonumber\\
	&& k_D(u_0) \equiv \frac{u_0\xi}{\hbar v} \sin(\phi+60^\circ).
\end{eqnarray}
In Fig.\ \ref{fig_twomode_3D}, 
we plot the energy dispersion Eq.\ (\ref{eq_E1D_two_modes_approx})
for the case of $\nu=1$ with black curves,
which can be recognized as the intersect between the shifted Dirac cones
and $k$-space quantization plane.

The energy band gap of DWNT
is determined by
the conduction band minimum of $E_1$
\begin{equation}
	E^{(c)} = -u_0 + \hbar v |m-m_D(u_0)|,
\end{equation}
and the valence band maximum of $E_2$
\begin{equation}
	E^{(v)} = u_0 - \hbar v |m+m_D(u_0)|.
\end{equation}
The difference
\begin{align}
	&\Delta E = E^{(c)} - E^{(v)}
	\nonumber\\
	&= -2u_0 + \hbar v |m-m_D(u_0)| + \hbar v |m+m_D(u_0)|
	\label{eq_energy_difference}
\end{align}
shows that
the DWNT can have a finite gap of
\begin{equation}
	\Delta E = 2(\hbar v |m|-u_0)
\end{equation}
only when $\hbar v |m| \ge u_0$.
Compared to the gap
in the absence of interlayer interaction, $2\hbar v|m|$,
we can see that
the interlayer interaction in DWNT
reduces the gap of the system
by $2u_0$ in a strong coupling condition.

Figure \ref{fig_twomode_1D}(a) 
shows the numerically calculated
band dispersions in the extend zone scheme
for (35,19)@(40,24) DWNT, 
plotted along the quantization line
closest to $\Vec{K}_\xi$. 
We can see a good consistency with
the analytic expression in Fig.\ \ref{fig_twomode_1D}(b),
which is calculated by Eq.\ (\ref{eq_E1D_two_modes_approx}).



Besides, since armchair-armchair DWNT is
one example of the strong coupling condition,
its energy dispersion [Fig.\ 3(a) in the main text]
\begin{eqnarray}
	&& E^\pm_1(k) = -u_0 \pm (\hbar vk + u_0\xi),
	\nonumber\\
	&& E^\pm_2(k) = u_0 \pm (\hbar vk - u_0\xi),
\end{eqnarray}
is also reproduced by
setting $\phi = 30^\circ$ and $\nu = 0$ in
Eq.\ (\ref{eq_E1D_two_modes_approx}).

\bibliography{dwnt}

\end{document}